\documentstyle[12pt]{article}
\newcommand{\be}{\begin{equation}}
\newcommand{\ee}{\end{equation}}
\newcommand{\bea}{\begin{eqnarray}}
\newcommand{\eea}{\end{eqnarray}}
\newcommand{\brr}{\begin{array}}
\newcommand{\err}{\end{array}}
\newcommand{\bc}{\begin{center}}
\newcommand{\ec}{\end{center}}

\begin{document}
\pagestyle{empty}
\begin{flushright}
ROME prep. 94/1042 \\
SHEP prep. 93/94-31 
\end{flushright}
\centerline{\Large{\bf{The Gluon Propagator}}}
\vskip 3mm
\centerline{\Large{\bf{on a Large Volume, at $\beta=6.0$.}}}
\vskip 1cm
\centerline{\bf{P. Marenzoni$^{a}$,  G. Martinelli$^{b}$, N. Stella$^{c}$}} 
\centerline{$^a$ Dip. di Ingegneria dell'Informazione}
\centerline{Universit\`a di Parma, Viale delle Scienze, 43100 Parma, Italy}
\centerline{$^b$ Dip. di Fisica,
Universit\`a degli Studi di Roma ``La Sapienza" and}
\centerline{INFN, Sezione di Roma, P.le A. Moro 2, 00185 Rome, Italy. }
\centerline{$^c$ Physics Department,
``The University'', S09 5NH Highfield, Southampton, U.K.}
\centerline{}
\centerline{}
\begin{abstract}
We present the results of a high statistics lattice study of the gluon 
propagator, in the Landau gauge, at $\beta=6.0$. As suggested by 
previous studies, we find that, in momentum space,
the propagator  is well described by the 
expression $G(k^2)= \Big[ M^2 + Z\cdot k^2(k^2/\Lambda^2)^\eta\Big]^{-1} $.

By comparing $G(k^2)$ on different volumes,
we obtain a precise determination of the exponent
$\eta=0.532(12)$, and verify that $M^2$ does not vanish
in the infinite volume limit.
The behaviour of $\eta$ and $M^2$ in the continuum limit is not known, and can 
only be studied by increasing the value of $\beta$.

\end{abstract}

\vspace{4cm}
\centerline{\em Submitted to Nuclear Physics B}

\newpage
\pagestyle{plain}
\setcounter{page}{1}

{\Large \bf Introduction}
\vskip 3mm 
\par\noindent
One of the most interesting, yet unresolved, issues concerning Quantum 
Cromodynamics, as a theory of strong interactions, is the comprehension
of the transition from the fundamental fields, appearing in the 
Lagrangian, to the physical particles, namely the hadrons.
While color confinement and hadronization are believed to occur, 
no consistent derivation of this mechanism from first principles is 
yet available, although much effort has been put in this direction.

By studying the propagation and the interaction of the 
fundamental fields,
appearing in the QCD Lagrangian, one attempts to gain some insight on the 
mechanism of confinement and hadronization, which are genuine non-perturbative 
effects. In this paper, we present the results of a non-perturbative study of the 
gluon propagator, in the Landau gauge, obtained from a high statistics lattice
simulation.

The results of the present study, on a volume larger than in refs.  
\cite{NOSTRO,PARRI}, 
confirm that the gluon  propagator is well described by
the functional form proposed in the references above:
\begin{equation}
G(k^2)= \left[M^2+Z\cdot(k^2)\left(
\frac{k^2}{\Lambda^2}\right)^\eta\right]^{-1}.
\end{equation}
We obtain a more
precise determination of the exponent, finding 
$\eta=0.532(12)$ and show that the gluon propagator at zero momentum stays
finite, as the infrared cut-off is removed.
 
In section 1, we review what is known from perturbation theory, 
and some results obtained by perturbative analyses. 
Particular attention is devoted to other lattice simulations, 
whose results can be compared directly with ours.
In section 2, the definition of the correlation functions 
will be given, together with a description of the numerical implementation.
Section 3 is dedicated to the discussion of systematic effects 
and lattice artefacts. We found important to discuss these aspects before 
giving the results, which are largely influenced by our control of systematic
errors.
Finally, in section 4, we present the new results and compare them 
with those of our previous study \cite{NOSTRO}, trying to interpret common 
features and differences.

\section{The gluon propagator in the literature}

As is the case for gauge-dependent quantities, 
the gluon propagator can only be defined 
once that the gauge has been fixed. In the following, we will solely be 
concerned with the Landau gauge, which corresponds to
fulfill the condition
\begin{equation}
\partial_{\mu} A_{\mu}^{a}(x)=0, \qquad a=1, \ldots ,8 ,
\label{gauge}
\end{equation}
where $A^{a}_{\mu}(x)$ are the gauge fields, and $a$ the color index.

The gluon propagator in this gauge has the form
\begin{equation}
D_{\mu\nu}(k)= -i \int d^4 x \langle |T \left( A_{\mu}(x) A_{\nu}(0) 
\right) |\rangle e^{ikx} =
G(k^2) \left( g_{\mu\nu} - \frac{k_{\mu}k_{\nu}}{k^2} \right),
\label{perturbativeM}
\end{equation}
where $G(k^2) = -1/\!k^2$ at tree level. When leading order perturbative 
corrections are taken into account, $G(k^2)$ takes the form
\begin{equation}
G(k^2) = -\frac{1}{k^2(1+\Pi(k^2))} \simeq -\frac{1}{(k^2)^{1+\eta}},
\end{equation}
with $\eta^{\rm pert}= - \gamma_G g^2/4\pi$ and $\gamma_G=-3/13$.
Therefore, at values of the coupling constant corresponding to $\beta=6.0$, 
$g^2\sim 1$, these corrections
correspond to a minor change, at most 5\%, to the tree level 
expression\footnote{ To obtain this extimate we have not considered the running of the 
coupling constant.}. 

Perturbative corrections to $G(k^2)$ cannot therefore account for the higher power of
$k^2$ which is necessary to 
give a confining potential, in the one gluon-exchange approximation.

An analytic non-perturbative approach to the problem is 
indeed very difficult.
The most elegant and neatest way to attack the problem is perhaps to
solve the Dyson-Schwinger equations w.r.t. the renormalized gluon two-point 
function $D_{\mu\nu}(k)$. The infinite set of differential equations should 
satisfy the Slavnov-Taylor identities in order to preserve gauge invariance. 
Being the equations non linear, different solutions are allowed
\cite{BP1}--\cite{STINGL},  and 
their interpretation, already difficult, is made even more uncertain by the 
approximations that one has to make to render the problem feasible.

In ref. \cite{GRIBOV} a procedure for deriving the gluon propagator in the 
Minimal Landau Gauge, is presented.
Subsequently, Zwanziger confirmed \cite{ZW} the solution of ref. \cite{GRIBOV} 
by repeating the calculation on the lattice. In ref. \cite{NOSTRO}, 
we have shown that this expression fails to reproduce the lattice behaviour.

In ref. \cite{MANDULA}, the first exploratory study on the lattice was presented.
They found that the effective energy is an increasing function of the time
(see below for definition), that the
gluon propagator violates the spectral decomposition and, in particular, 
cannot have the form (\ref{perturbativeM})  with $G(k^2)\sim 1\!/(k^2+m^2)$.
Subsequently, the authors of ref.  \cite{GUPTA,PARRIold} tried to account for 
such a behaviour either 
by including the possible contribution of unphysical particles, 
or considering the solution proposed in refs. \cite{GRIBOV,ZW}.

Lately, to interpret the results of the lattice simulations, a new functional 
form has been proposed \cite{NOSTRO,PARRI}:
\begin{equation}
G(k^2)= \left[M^2+Z\cdot(k^2)\left(\frac{k^2}{\Lambda^2}\right)^\eta
\right]^{-1}.
\label{ourfunc}
\end{equation}
This expression tries to mimic the effect of an anomalous dimension
for the gauge field, and keeps $G(k^2)$ finite. 
The latter is a necessary condition, due to the fact that the lattice possesses an
intrinsic infrared cut-off.

According to \cite{NOSTRO} $\eta=0.4 \div 0.6$, whereas the authors of \cite{PARRI} 
quote $\eta=0.35$. The value of the dimensionful parameter $M^2$ is 
more controversial, and difficult to determine because
it is very small, of the order of the infrared cut-off. It is not clear 
whether $M^2$ will vanish in the infinite volume limit.

Since there is no a priori theoretical prediction of the functional 
form (\ref{ourfunc}),
several further investigations are in order. Namely,
\begin{enumerate}
\item
reduce the ultraviolet cut-off by increasing the value of $\beta$, and
try to determine $M^2$ and $\eta$ in the continuum limit;
\item
verify if (\ref{ourfunc}) contains any gauge-independent information.
This can be investigated by studying the gluon propagator in different gauges.

\end{enumerate}

In this paper, we address a different issue, that is the dependence of the parameters
of equation (\ref{ourfunc}) on the infrared cut-off. Anticipating some of the results,
we have indications that the dimensionful
parameter $M^2$  in not merely a finite volume artefact, and it would keep a 
non-zero value in the infinite volume limit. Its fate in the continuum limit, 
i.e. as $\beta \to \infty $, is not known.
The value of the anomalous dimension $\eta$,
on the other hand, does not seem to be affected by the presence of the infrared regulator.
A major question is the interpretation of a finite value
of the gluon propagator, at zero momentum.

\section{The gluon propagator on the lattice}
In this section, we establish the notation and the main formulae, 
to be used in the following.
We also give a brief description of the algorithms, i.e. the Monte Carlo 
procedure, used to generate statistically independent configurations, 
and the method which has been used to fix 
the Landau gauge. We conclude by reporting some information about 
the accuracy achieved by our gauge-fixing algorithm.

\subsection{Notations}
The gluon propagator is studied on the lattice by looking at the euclidean 
two-point correlation functions:
\begin{equation}
D_{\mu\nu}(x,y) = {\rm Tr}\langle A_{\mu}(x) A_{\nu}(y)\rangle,
\label{2pt} 
\end{equation}
where $\mu,\nu = 1,\ldots ,4 $ and the trace is intended over color indices.
The corresponding expression in momentum space is
\begin{equation}
D_{\mu\nu}(k)= G(k^2) \left( \delta_{\mu\nu} - \frac{k_{\mu}k_{\nu}}{k^2} 
\right).
\label{perturbative}
\end{equation}

Among the possible definitions of the gluon field on the lattice, we choose
\begin{equation}
A_{\mu}(x) = \frac{U_{\mu}(x) - U_{\mu}^{\dag}(x)}{2i}.
\end{equation}
In a previous publication \cite{NOSTRO}, we have discussed in detail the 
motivations for such choice.

In order to study the gluon propagator in momentum space, $D_{\mu\nu}(k)$, 
we take the 4-dimensional Fourier transform of the correlators (\ref{2pt}). 
Taking into account translation invariance, $D_{\mu\nu}$ is given by
\begin{equation}
D_{\mu\nu}(k) = \sum_{\vec{x},t}{\rm Tr} 
\langle A_{\mu}(x) A_{\nu}(0)\rangle e^{ikx}.
\label{mom-2pt}
\end{equation}
On the lattice, $k$ can only take discrete values, 
$k_{\mu}=(2\pi\!/L_{\mu} a)n$, where $a$ is the lattice spacing,
and $n=0, 1,\ldots ,L_{\mu} -1$.

Moreover:
\begin{itemize}
\item
we always take full advantage of space-time translational symmetry by averaging 
the previous expression over different points on the euclidean lattice;
\item
from r.h.s.\ of (\ref{mom-2pt}) the scalar function $G(k^2)$ is extracted,
by taking suitable combinations of the spin indices $\mu, \nu$ accordingly to
(\ref{perturbative}). 
\item
in order to check the validity of continuum dispersion relation on the lattice,
the following four classes of correlators have been considered:
\begin{enumerate}
\item
\underline{Transverse correlations}
\begin{equation}
D_{ii}(k) = \sum_{\vec{x},t}{\rm Tr} 
\langle A_{i}(x) A_{i}(0)\rangle e^{ i k_j x_j}e^{ik_0 t}=
G(k^2) \delta_{ii},
\label{trans_k}
\end{equation}
where $i,j=1,\ldots ,3$ and the spatial momentum $k_j$ is non-zero only along axis
orthogonal to the spin direction $i$. 
\item
\underline{Longitudinal correlations}
\begin{equation}
D_{jj}(k) = \sum_{\vec{x},t}{\rm Tr} 
\langle A_{j}(x) A_{j}(0)\rangle e^{ i k_j x_j}e^{ik_0 t}
=G(k^2) \left( \delta_{jj} -\frac{k_j^2}{k^2} \right),
\label{long_k}
\end{equation}
where $j=1,\ldots ,3$ and the spatial momentum has a non-zero component along 
the direction of the spin $j$. 
\item
\underline{Non-diagonal correlations}
\begin{equation}
D_{ij}(k) = \sum_{\vec{x},t}{\rm Tr} 
\langle A_{i}(x) A_{j}(0)\rangle e^{ i \vec{k}\cdot \vec{x}}e^{ik_0 t}
= - G(k^2) \frac{k_i k_j}{k^2},
\label{nond_k}
\end{equation}
where $i,j=1,\ldots ,3$ and $i\neq j$. These correlations  do not vanish 
only when the spatial momentum is non-zero along both $i$ and $j$ directions.
\item
\underline{Temporal correlations}
\begin{equation}
D_{00}(k) = \sum_{\vec{x},t}{\rm Tr} 
\langle A_{0}(x) A_{0}(0)\rangle e^{ i \vec{k}\cdot \vec{x}}e^{ik_0 t}
= G(k^2) \left( \delta_{00} -\frac{k_0^2}{k^2} \right).
\label{time_k}
\end{equation}
\end{enumerate}
\end{itemize}

Further investigations of the behaviour of the gluon propagator can be
 carried on by looking at some correlations, which are functions of the 
spatial momentum and euclidean time. For simplicity we maintain
the same notation given before, namely, we will consider:
\begin{enumerate}
\item
\underline{Transverse correlations}
\begin{equation}
D_{ii}(t,\vec{k}) = \sum_{\vec{x}}{\rm Tr} 
\langle A_{i}(x) A_{i}(0)\rangle e^{ i k_j x_j},
\label{trans_t}
\end{equation}
\item
\underline{Temporal correlations}
\begin{equation}
D_{00}(t,\vec{k}) = \sum_{\vec{x}}{\rm Tr} 
\langle A_{0}(x) A_{0}(0)\rangle e^{ i \vec{k}\cdot \vec{x}}.
\label{time_t}
\end{equation}
\end{enumerate}

All the previous expressions have been computed for the following values of 
spatial momenta:  $\vec{k}=(0,0,0), (2\pi\!/24a,0,0), (4\pi\!/24a,0,0)$,  and
$\vec{k}=(2\pi\!/24a,2\pi\!/24a,0) $. Moreover, permutations of the three
spatial axes have been considered. The Fourier transform, with respect to the 
time coordinate has been computed allowing $k_0$ to take all the possible 
values. Namely, $k_0=( 2\pi\!/48a)n,$ with $n=0,\ldots ,47$ .

\subsection{Lattice setup and gauge-fixing}
We summarize the parameters of the present simulation in tab. \ref{tab.setup}.
The only differences, with respect to ref. \cite{NOSTRO}, are the lattice size
and the number of configurations. All the others 
parameters of the simulation have been left unchanged,
with the purpose of studying the dependence of expression 
(\ref{ourfunc}) on each of them.
For the generation of the pure gauge fields we adopted a Hybrid
Monte Carlo (HMC) algorithm in its local version
and with a checkerboard update, as described in \cite{LHMC}.
This version of the HMC allows very short autocorrelation times
among pure gauge configurations and is very effective when 
high statistic measures are needed.
In our simulation we performed an initial termalisation of 1024 iterations
and then we extracted configurations every 128 iterations.

A further comment on the size of the statistical sample:
although the number of configurations
generated during the present run, is a half the old one, we have obtained more 
accurate results.
This is due to the fact that the larger number of lattice 
points (the ratio of the two volumes is about 8) improves the stability of the
signal.

\begin{table}
\begin{center}
 \begin{tabular}{|l|c|c|c|c|}\hline
                   & $\beta$ & $\#$ confs. & Volume         & $\partial_{\mu} A_{\mu}$ \\ 
\hline\hline
present work       & 6.0     &    500   & $24^3\times 48$& $ < 10^{-6} $ \\
\hline
ref. \cite{NOSTRO} & 6.0     &   1000   & $16^3\times 32$& $ < 10^{-6} $ \\
\hline
  \end{tabular}
\end{center}
\caption{Summary of the parameters of the present simulation and of those of the 
simulation
of ref. [1]}
\label{tab.setup}
\end{table}

In presence of gauge-dependent quantities, the accuracy of the
gauge-fixing algorithm is a crucial information. Uncertainties in the gauge-fixing
result in residual fluctuations,
thus causing extra noise and/or systematic alterations. 
For this reason, the accuracy can only be established by 
performing tests on the gauge-dependent operators under study.

As demonstrated in ref. \cite{NOSTRO},
the residual fluctuation left-over after gauge-fixing 
\begin{equation}
\langle\partial_{\mu} A_{\mu}(x)\rangle_{{\rm Latt}} \le 10^{-6},
\label{precision}
\end{equation}
are absolutely negligible, with respect to the statistical errors. Since we consider 
this to be a crucial point, we have repeated the same analysis as in \cite{NOSTRO}
on the new sample.
We have defined logaritmic (symmetric)
derivatives, with respect to $t$,
of transverse and temporal correlators, at zero spatial momentum. 
As in ref. \cite{NOSTRO}, we find that the temporal correlator is constant and its derivative always compatible 
with zero, within errors. 
This qualitative behaviour can be compared with a similar figure in 
ref. \cite{GUPTA}.

A different check can be performed by noting that the condition (\ref{gauge}) implies that the three dimensional 
Fourier transform of the gauge field $A_0(t,\vec{k}=\vec{0})$ 
is time independent. This time independence should hold 
on each single configuration.
This has been checked by looking at the temporal correlation
$D_{00}(t,\vec{0})$ of some arbitrarily
chosen configurations. In table 
\ref{tab.gauge}, we report the value of $D_{00}(t,\vec{0})$, at different time 
separations, on a single configuration. $D_{00}(t,\vec{0})$ is constant at the level of 
$ 0.008\% $, which is in fact consistent with the precision (\ref{precision}).
Results in table \ref{tab.gauge} are to be compared with figure 1 of ref.
\cite{PARRI}, where the gluon field, plotted versus $t$
after gauge-fixing, shows a strong time dependence.

\begin{table}
\begin{center}
 \begin{tabular}{|rr|rr|rr|}\hline
$t$ & $D_{00}(t,\vec{0})$ & $t$ & $D_{00}(t,\vec{0})$ & $t$ & $D_{00}(t,\vec{0})$ \\ 
\hline
1   & 56552 & 6  & 56547 & 11 & 56545 \\
2   & 56552 & 7  & 56549 & 12 & 56547 \\
3   & 56550 & 8  & 56548 & 13 & 56542 \\
4   & 56549 & 9  & 56547 & 14 & 56543 \\
5   & 56548 & 10 & 56548 & 15 & 56543 \\
\hline
\end{tabular}
\end{center}
\caption{ $D_{00}(t,\vec{0})$, measured on one configuration, 
as a function of the time separation.
The gluon field $A_{0}(t,\vec{k})$ is constant in time, as required by
the gauge condition.}
\label{tab.gauge}
\end{table}

The previous discussion shows that the present gauge-fixing procedure is 
the most effective one, among those implemented in the literature.

\section{Results for the gluon propagator}
In this section, we outline  the unusual
behaviour of the effective energy.
Then, we consider the dependence of $G(k^2)$ on the momentum.
Assuming the validity of the functional form (\ref{ourfunc}), 
we first discuss in length systematic effects, and give our final results. 

\subsection{The behaviour of the effective energy}
Using spectral decomposition and translation invariance, 
the correlation functions  can be rewritten as
\begin{equation}
D(t,\vec{k}) = \sum_{|i\rangle}
\frac{\left| \langle A(0)|i\rangle \right|^2 }{{\cal N}_i} e^{ -E_i t},
\label{spectral}
\end{equation}
where the sum is intended over all the states with 
the quantum numbers of operator $A(x)$ and $E_i=\sqrt{m_i^2 + |\vec{k}|^2}$ is the energy 
of the state $i$. The normalization ${\cal N}_i$ is positive for physical states.
From eq. (\ref{spectral}), the effective energy 
\begin{equation}
\omega_{{\rm eff}}(t,\vec{k}) = \log \frac{D(t,\vec{k})}{D(t+1,\vec{k})}.
\end{equation}
should be a decreasing function of the time, 
for any value of the momentum $\vec{k}$.

\begin{figure}   
  \begin{picture}(90,100)(-18,-10)   
      \put(-10,-40){\special{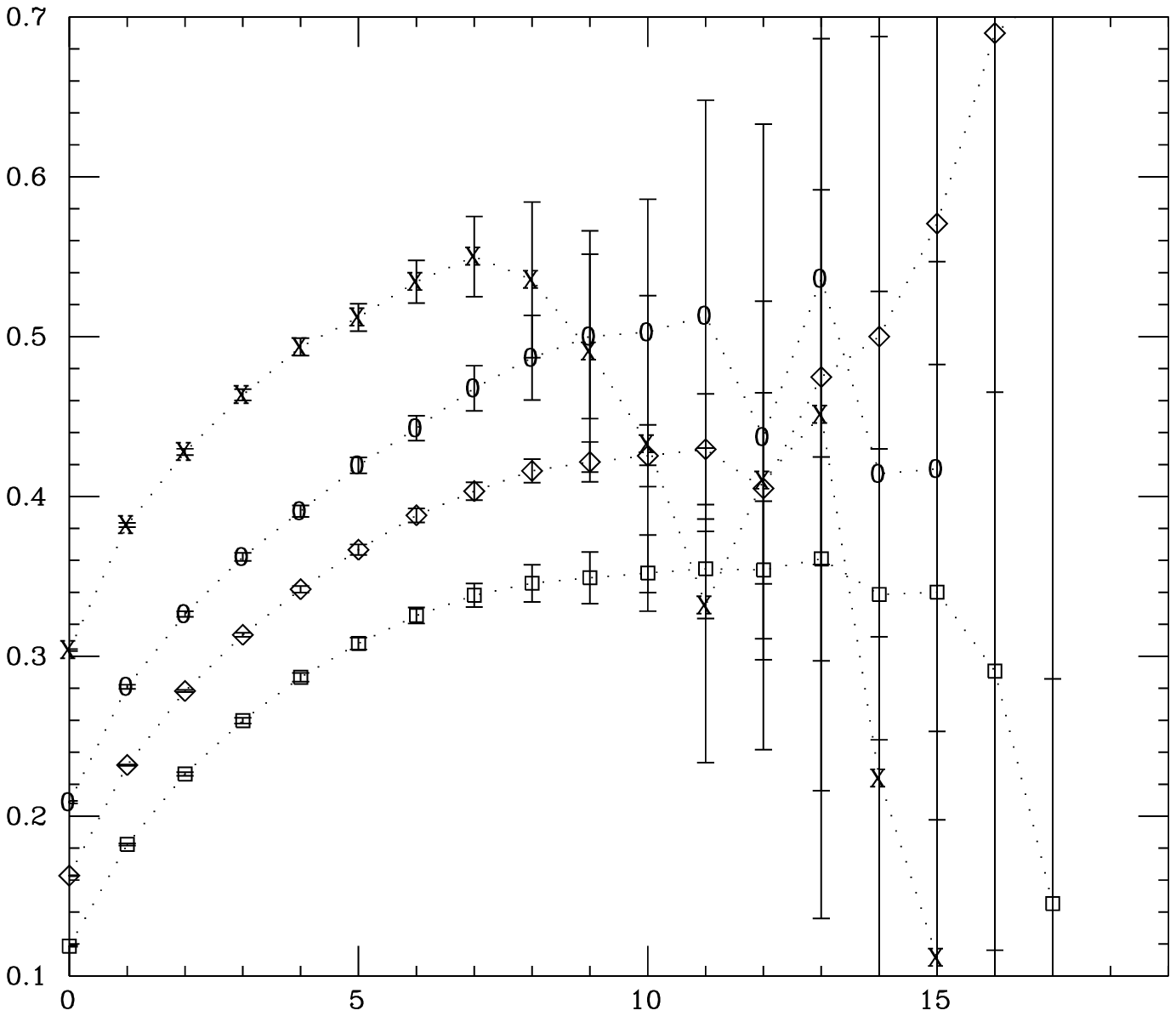}}   
      \put(100,-10){$t$}
      \put(-17,80){$\omega_{\rm eff}(t,\vec{k})$}
  \end{picture}
 \caption{Effective energy for transverse correlators (20).
 The effective energy is  increasing with  time for all the values of 
$\vec{k}$. The four curves correspond to the following momenta: 
$\Box :\vec{k}=\vec{0}$, 
$\Diamond :\vec{k}=(2\pi\!/24,0,0)$, x$: \vec{k}=(2\pi\!/24,2\pi\!/24,0)$
and $0 :\vec{k}=(4\pi\!/24,0,0)$}
 \label{fig.mtutte2}
\end{figure}

In figure \ref{fig.mtutte2}, we report $\omega_{\rm eff}(t,\vec{k})$ computed
from the transverse correlators (\ref{trans_t}). $\omega_{\rm eff}(t,\vec{k})$
is  clearly increasing with time, for all the momentum combinations considered.
Some authors \cite{GUPTA} tried to justify this result by including 
(\ref{spectral})  the contribution of unphysical particles, whose normalization is negative.
The function apt to fit the lattice points, at large time separations, would then 
be:
\begin{equation}
F(t) =
\frac{Z_1}{2m_1}e^{-m_1t} + \frac{Z_2}{2m_2}e^{-m_2t};  \qquad Z_1 < 0, Z_2>0;
  \qquad m_1>m_2.
\label{2ma}
\end{equation}
In our previous paper \cite{NOSTRO}, we showed that (\ref{2ma}) fails to 
reproduce the behaviour of $D_{\mu\nu}(t,\vec{k})$. 
The apparent agreement of 
(\ref{2ma}) with $D(t,\vec{k}=\vec{0})$ in ref. 
\cite{GUPTA}, was probably faked by the large statistical errors. 

\subsection{Systematic effects}
The systematic effects discussed in the following
are those due to the infrared and ultraviolet cut-offs, introduced 
when the theory is regularized on a finite lattice. These effects limit
the range of momenta, where reliable  results can be 
obtained. A way to detect such effects
is to look for discrepancies between continuum dispersion relations
and their lattice realization.
We note that, contrary to what happens for other quantities, the quenched approximation
is not going to produce systematic changes, being the pure gauge sector a 
self-consistent part of the theory. One then expects these results to stay valid
{\em per se}, as a description of the pure gauge behaviour, although they might
differ in the full theory.
Another source of systematic uncertainty is the value of lattice spacing,
which is found to vary substantially depending on the physical 
observable used to fix it. We will only be concerned about this point at the
end of this section, when we will express, in physical unit, 
the propagator at zero momentum.

\begin{figure}   
  \begin{picture}(90,80)(-20,-5)   
      \put(-20,-55){\special{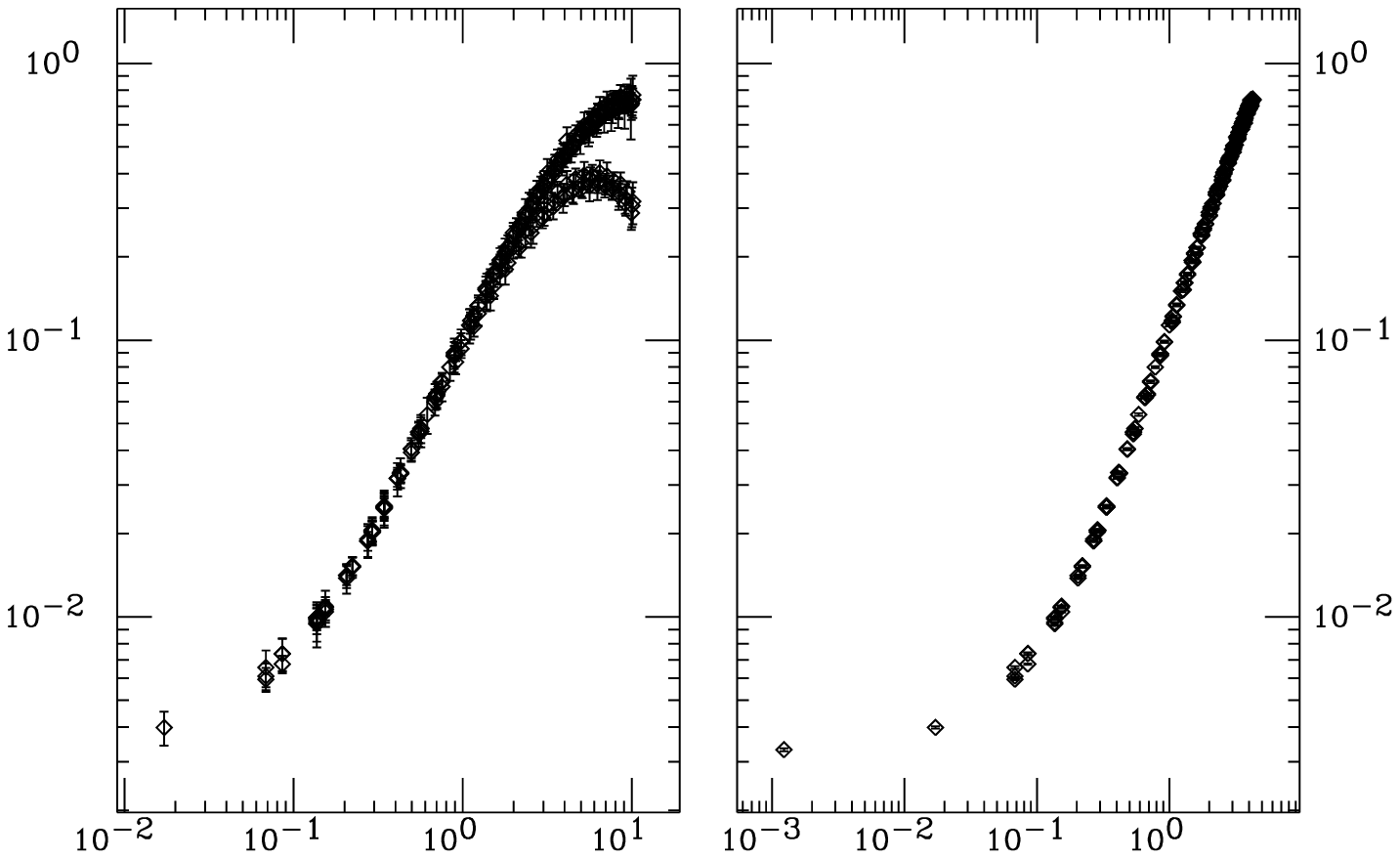}}   
      \put(0,60){Figure A}
      \put(70,60){Figure B}
      \put(32,-5){$k^2[a^2]$}
      \put(95,-5){${\cal K}^2[a^2]$}
      \put(-22,65){$G^{-1}(k^2)$}
  \end{picture}
 \caption{Behaviour of $G^{-1}(k^2)$ at high momenta. Fig. 2A: we observe 
the data splitting
into two curves at high momenta ($k^2a^2\sim 1$). Fig. 2B:
the two curves are mapped into a single line when the kinematical
correction (26) is applied. The logarithmic behaviour of the corrected 
curve is much closer to a straight line.}
 \label{fig.hmom}
\end{figure}

We discuss the effects of the ultraviolet cut-off, first. It is known that
the presence of a finite ultraviolet cut-off produces unwanted effects, 
when momenta are of the order of the cut-off itself. On the lattice, this means 
that results, corresponding at momenta such that $k^2 a^2 > 1$, 
contain discretization contributions.
A demonstration of the above statement, is given in figure \ref{fig.hmom}A,
where $G(k^2)$ is plotted as a function of
$k^2$ in dimensionless units. In correspondence of the same value of $k^2$,
$G(k^2)$ can have different values, due to the breaking of the Lorentz
symmetry. The different results correspond to $G(k^2)$,
as extracted from the four classes of correlators 
(\ref{trans_k})--(\ref{time_k}). 
They agree within errors until $k^2 a^2\simeq 1$ where
they start clustering into two separated curves.
This behaviour, at larger momenta, is caused by the fact that, in the Fourier transforms 
(\ref{trans_k})--(\ref{time_k}),
we only increase
$k_0$, to the larger values allowed on the lattice, while $k_i, i=1,2,3$ are
always kept small. 
Points lying on the higher curve are those generated through spatial correlators
(\ref{trans_k})--(\ref{nond_k}), while, on the lower curve, there are the points 
computed from the temporal correlators (\ref{time_k}), which are directly proportional
to $k_0^2$ and so more affected by discretization effects, at large $k_0$.

To verify this hypothesis, we have tried a kinematical correction to the momentum, 
substituting
\begin{eqnarray}
k_{\mu} &\longrightarrow & {\cal K}_{\mu}= 2 \sin(\frac{k_{\mu}}{2a}), \nonumber\\
k^2     &\longrightarrow & {\cal K}^2 = \sum_{\mu=1}^4 \big( 2 \sin(\frac{k_{\mu}}{2a})\big)^2 
\label{corre}
\end{eqnarray}
in the dispersion relation (\ref{perturbative}). Such a correction is 
suggested by the tree level expression of the bosonic propagator 
on the lattice. 
In fig. \ref{fig.hmom}B, $G(k^2)$, obtained by performing the substitutions
(\ref{corre}), is shown. One sees that the two curves of fig. \ref{fig.hmom}A
have collapsed into a single one. On a logarithmic scale, the curve, obtained by 
using (\ref{corre}) appears
as a straight line, at large $k^2$, whose slope
is the exponent $1+\eta$ of equation (\ref{ourfunc}). Around 
${\cal K}^2a^2=1$ the curve changes slope and $\eta$ is slightly
reduced. We cannot tell whether this behaviour is due to residual cut-off effects,
or to the fact that, as the momentum increases, $\eta$ is going towards its 
perturbative value. 
This issue can only be decided by decreasing the lattice spacing, i.e. by going to
larger values of $\beta$,
and increasing the range of $k^2$, where lattice artefacts are small and momenta are
large in physical units.

The investigation of the effects of the infrared cut-off is delicate and
 is directly related to the determination of the value of $M^2$ in (
\ref{ourfunc}).
At very low momenta, different determinations of $G(k^2)$ corresponding to 
the same value of $k^2$ are inconsistent (see fig. 3). This effect
is due to the fact that our lattice has  different length, along
the spatial and temporal axes. In fig. 3, we indicate, with an arrow, the
value of $k^2$ where the differencies became negligible.

We deduce from the previous analysis that one should only take into consideration 
the region of of intermediate momenta. More quantitative 
indications in this sense are provided the fitting procedure, see below.

\subsection{Fitting procedure}
In the previous section, it has been shown that  infrared and ultraviolet effects are reasonably small,
 only in the intermediate 
region of momenta.
On large volumes, the region of momenta ``free'' from cut-off effects
contains many points, which can be used to check the stability of the fits.
The fitting function is the dimensionless version of (\ref{ourfunc})
\begin{equation}
G(k^2)= \frac{1}{M^2_{{\rm L}}  + Z_{{\rm L}}(k^2)^{1+\eta} } ,
\label{fitfunc}
\end{equation}
which depends on three free parameters. We fit first $G(k^2)$ to 
expression (\ref{fitfunc}), on an interval of 6 
points, which we move back and forth in the selected region of momenta. 
The fit is then repeated using the same procedure with 8 and 12 points,
respectively. In the central region of $k^2$, 
the results are perfectly compatible 
on different $k^2$-ranges and their stability improves, by increasing the 
number of points. 

In figure \ref{fig.3fit}, we plot the best 12-points 
fits for $\eta$ and $M^2$, together with the corresponding value of $\chi^2_{{\rm ndof}}$,
as a function of the minimum momentum considered in the fit ($k^2_{{\rm min}}$). 
As anticipated in the previous section, for low and high
momenta the fits are rather poor, giving unstable values for the parameters 
and a large  $\chi^2_{{\rm ndof}}$. 
At momenta $k^2_{{\rm min}}a^2\simeq 0.1$ 
the values of $\chi^2_{{\rm ndof}}$ lower to $1\div 1.5$ and the estimates 
for $\eta$ and $M^2$ become very stable.
Our final determination obtained by averaging over the four points 
indicated in the
figure, is
\begin{equation}
\left\{
\begin{array}{ccl}
M_L^2 &=& 4.46(9)\times 10^{-3} \\
Z_L &=& 0.102(1) \\
\eta &=& 0.532(12)  \\
\langle \chi^2_{{\rm ndof}}\rangle &=& 1.08 \ \  
\end{array}
\right.
\label{quote} 
\end{equation}
One can compare the present results with those of ref. \cite{NOSTRO}. 
We have reanalyzed the old result, by fitting $G_1(t)$ to expression 
(\ref{fitfunc})\footnote{In is necessary to change the overall normalization of 
$G_1(t)$, in order to make it
independent on the lattice volume.}.
On a volume $V=16^3\times 32$
we find:
\begin{equation}
\left\{
\begin{array}{ccl}
M_L^2 &=& 2.8(1)\times 10^{-3} \\
Z_L &=& 9.01(4)\times 10^{-2}\\\
\eta &=& 0.56(6)   \\
\chi^2_{{\rm ndof}} &=& 1.5 \ \  
\end{array}
\right.
\label{quoteold} 
\end{equation}

\section{Discussion of the results}
Apart from the value of $M^2$ (see next section), 
the agreement between the two
sets of results is reasonably good, and the quality of the fits is satisfactory.
By increasing the volume, we have obtained a more accurate estimate of the 
exponent $\eta$.

\begin{figure}   
  \begin{picture}(90,80)(-10,-5)   
      \put(-18,-55){\special{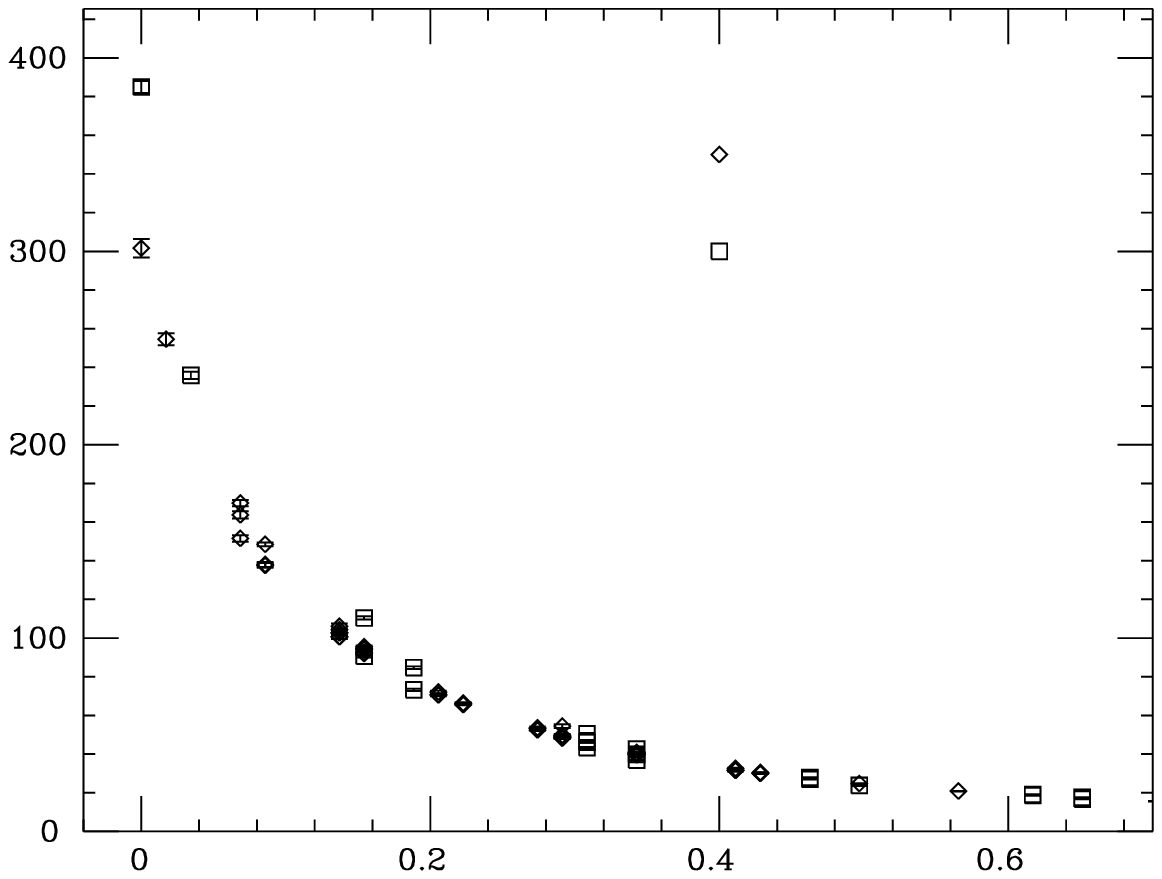}}   
      \put(82,72){$V=16^3\times 32$}
      \put(82,62){$V=24^3\times 48$}
      \put(82,-3){$k^2[a^2]$}
      \put(-2,72){$G(k^2)$}
       \put(39.5,38){\vector(0,-1){7}}      
  \end{picture}
 \caption{Behaviour of $G(k^2)$ at low momenta. 
The arrow indicates the value of $k^2$ where different estimates of $G(k^2)$ 
start to agree, within the errors.}
 \label{fig.lmom}
\end{figure}

In our previous paper, we have quoted three different sets of results, 
obtained by studying different 
correlators, as functions of $k^2$ or  $\vec{k},t$ respectively. 
We interpreted the differences as systematic effects. We are now
able to check this interpretation. 
On a smaller volume, we had a limited choice of momenta, before running into 
problems when $k^2a^2\simeq 1$ or infrared effects become important.
As figure \ref{fig.lmom} shows, some of the points 
we had considered in ref. \cite{NOSTRO}, in the fit of the propagator 
in momentum space, (see eq. (19) in ref. \cite{NOSTRO}), 
correspond to incompatible estimates of $G(k^2)$. 
For this reason, in ref. \cite{NOSTRO}, the fit performed in momentum space,
was giving results uncompatible with those of eqs. (22) and (23),
and it also disagrees with the present result (\ref{quote}).

In ref. \cite{PARRI}, all the values of momenta, (including those corresponding
to $k^2a^2\geq 1$) have been considered in the fit. Due to the flattening of the 
curve in the region of large momenta, 
the value of $\eta$ has been underestimated.

Finally, we observe that the finite value of $V$ has a very small effect in the 
determination of  $\eta$, provided that the range of $k^2$ is large 
enough.

\subsection{Comments on the value of $M^2$ }
The main difference between (\ref{quote}) and (\ref{quoteold}) is the value 
on the parameter $M^2$, which increases with the volume. This observation
rules out the hypothesis that the non-zero value found for $M^2$ is merely due
to finite volume effects. If this were the case, we
would expect $M^2$ to scale roughly as $1\!/L^2$.

Further arguments may be provided to reinforce the conclusion that $M^2\neq 0$.
In figure \ref{fig.lmom}, values of $G(k^2)$ measured on a $16^3\times 32$ lattice
are over-imposed to the present data. We notice that $G(k^2=0)$, whose inverse 
gives an estimate of $M^2$, is smaller on the largest volume. This is opposite
to the expectation $M^2 \propto 1\!/L^2$.

\begin{figure}   
  \begin{picture}(90,80)(-5,-5)   
      \put(-10,-55){\special{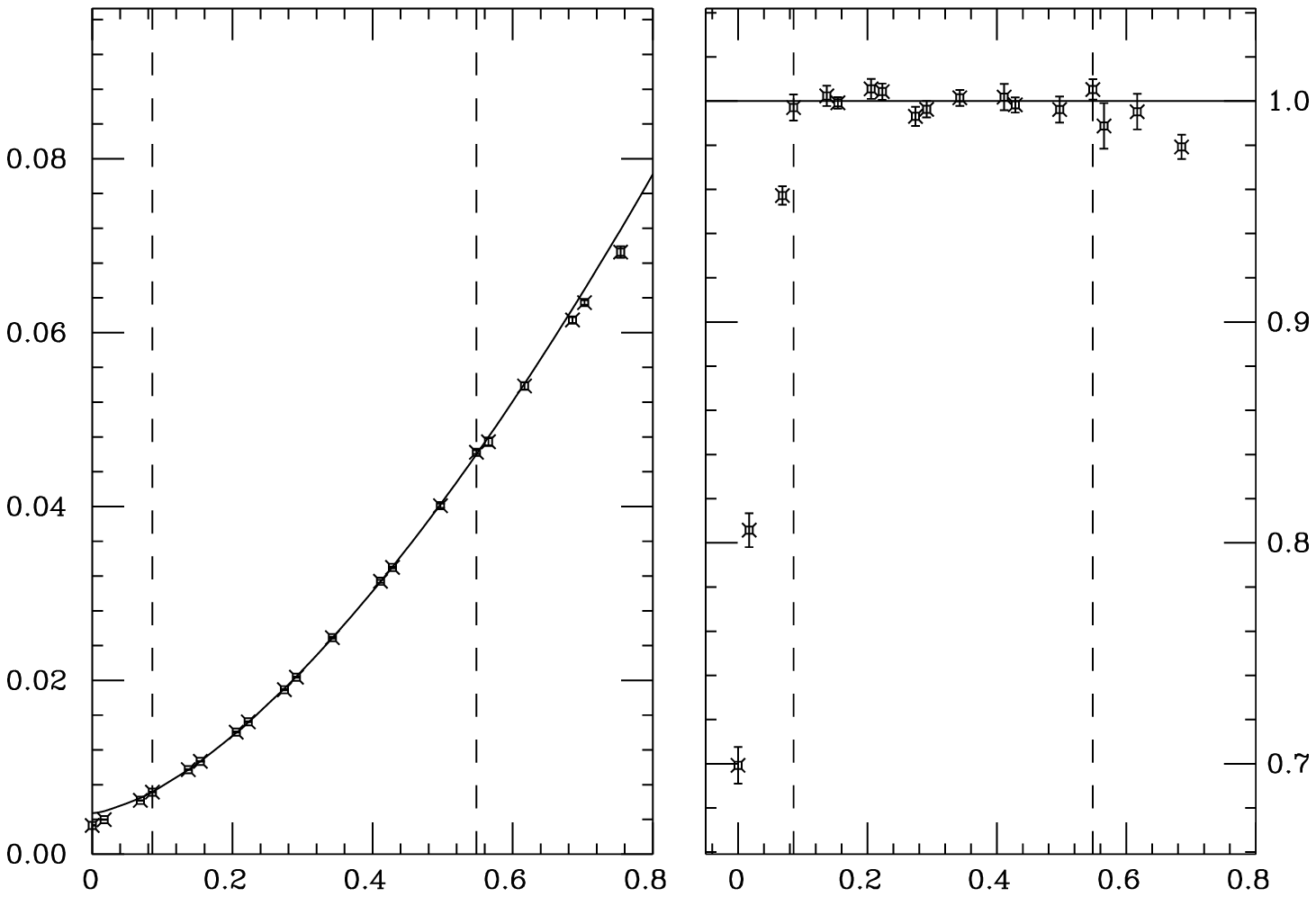}}   
      \put(-14,88){$G^{-1}(k^2)$}
      \put(50,-4){$k^2[a^{2}]$}
      \put(120,-4){$k^2[a^{2}]$}
  \end{picture}
 \caption{Best fit for the propagator in momentum space with the function
(27).
The figures show the fit (left) and the data to fit ratio, the dashed lines indicate
the fitting range. The ratio shows that the fit overshoots the data for small
momenta.}
 \label{fig.ratio}
\end{figure}

\begin{figure}   
  \begin{picture}(90,110)(-5,-7)   
      \put(100,107){$\chi_{{\rm ndof}}$}
      \put(100,67){$\eta$}
      \put(100,30){$M^2$}
      \put(75,-6){$k^2_{{\rm min}}[a^{2}]$} 
      \put(-10,-40){\special{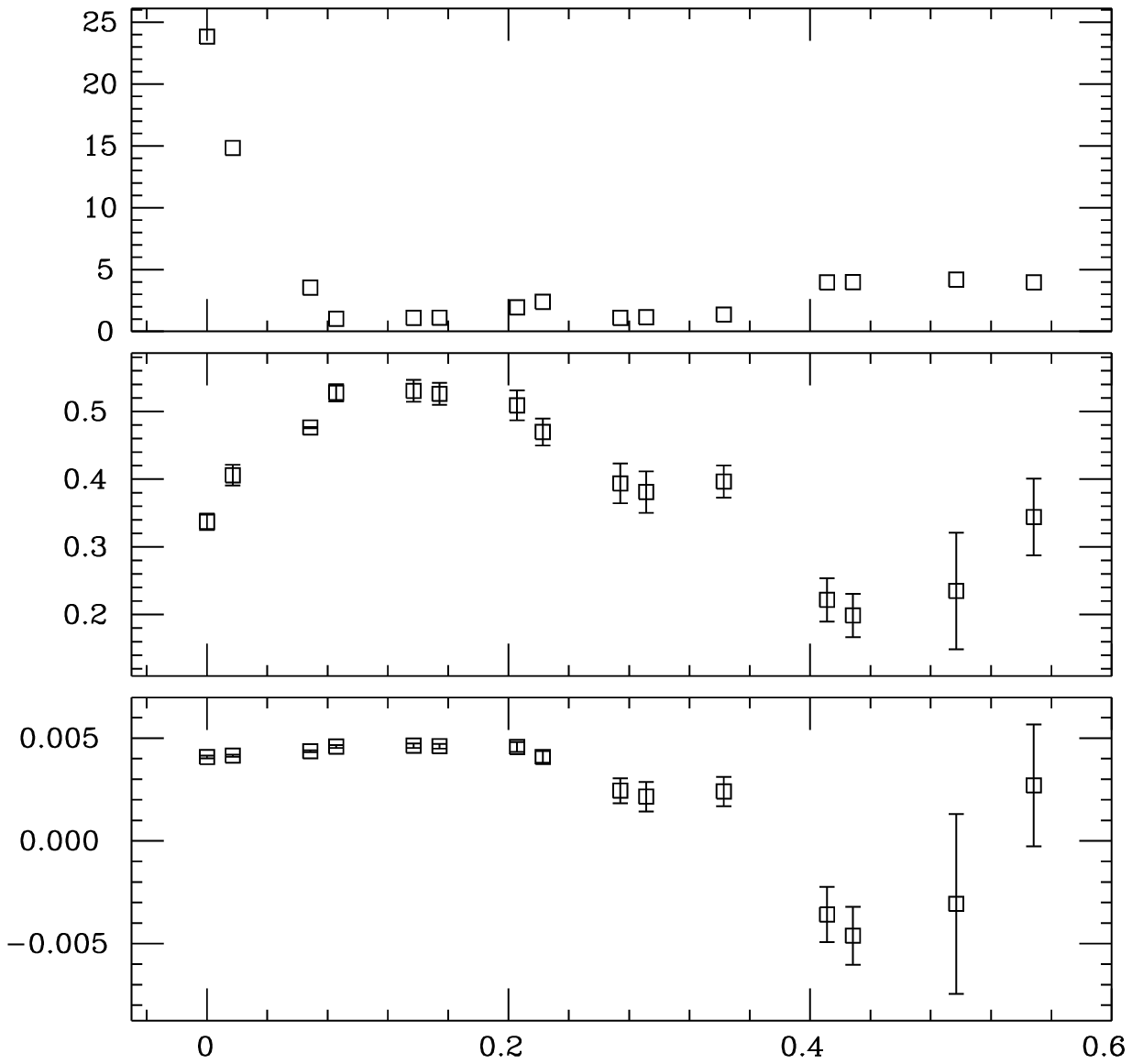}}   
      \put(44,22){\vector(0,1){7}}
      \put(52.5,22){\vector(0,1){7}}
      \put(55.5,22){\vector(0,1){7}}
      \put(64,22){\vector(0,1){7}}
      \put(44,61){\vector(0,1){7}}
      \put(52.5,61){\vector(0,1){7}}
      \put(55.5,61){\vector(0,1){7}}
      \put(64,61){\vector(0,1){7}}
  \end{picture}
 \caption{Values for $M^2$ and $\eta$ versus the minimum momentum
considered. The correspondent $\chi^2_{{\rm ndof}}$ is also shown.
In this fit we considered 12 points.}
\label{fig.3fit}
\end{figure}

A second argument is provided by the comparison of the fit to the data.
In figure \ref{fig.ratio}, the best fitting curve is compared to the 
lattice numbers, showing  a good agreement even outside the fitting range.
On the right side, the ratio data to fit 
shows that the fit overshoots the data, at low momenta. That is,
the behaviour of the propagator at the so called intermediate momenta is
consistent with a value of $M^2$ larger than $G(k^2=0)$ found on the lattice.
This is a further proof that the effect of the infrared cut-off is to 
fake a value of $M^2$ lower that the real one.

It is possible to try a first, very crude extrapolation of the value 
of $M^2$ to the infinite volume limit. Using 
\begin{equation}
M^2(V) =  M^2(V=\infty) + {\rm cost}\frac{1}{\sqrt{V}} ,
\end{equation}
we get
\begin{equation}
M^2(V=\infty) = 6.202(8)\times 10^{-3}.
\end{equation}

By using $a^{-1}\sim 2 $ GeV, as it is determined by several lattice simulations
at $\beta=6.0$, this corresponds to
\begin{equation}
M^2_{{\rm phys}} \simeq (160\ {\rm MeV})^2 \simeq (\Lambda_{QCD})^2
\end{equation}

It is fundamental to understand the behaviour of $M^2$ in the continuum, 
i.e. as 
$\beta \to \infty$. Notice that, even if $M^2$ stays finite in this limit, it 
cannot be interpreted as a symptom of symmetry breaking, since, in
 (\ref{ourfunc}), $M^2$ does not represent a pole.
An interpretation of this result, in connection with colour 
confinement is, at present, absent.

\section{Conclusions}
The study of the gluon propagator on different lattice volumes at fixed value of the 
coupling constant, confirms that $G(k^2)$
is well described by the expression (\ref{ourfunc}). The value of the anomalous dimension
$\eta$ has been confirmed to assume a value slightly exceeding $0.5$. 
By increasing the volume from $16^3\times 32$ to $24^3\times 48$, we have obtained a better 
determination of the anomalous dimension: $\eta = 0.532(12)$.
We find that $M^2$ does not vanish as $1\!/L^2$. Assuming $a^{-1}\simeq 2$ GeV,
our best estimate is $ M^2_{{\rm phys}} \simeq (160 {\rm MeV})^2 $.

Further investigations to study the continuum limit  and the 
gauge-dependence of the results presented above are certainly required. 
If, ultimately, $M^2$ should stay finite, 
the comprehension of the mechanism which allows the gluon
to develop a mass-scale in the infrared region would call for 
a theoretical understanding.

\section*{Acknowledgements}
We are largely indebted to M. Testa for valuable discussions and 
for the support he gave to this project.
We warmly thank the Thinking Machines Corporation for allowing us to
exploit the considerable amount of CPU-time and of memory, needed
for the present simulation. We are especially grateful to the staff of the
``Centre National de Calcule Parallel en Sciences de la Terre'' of Paris,
for their precious help during all the stages of the 
simulation. We acknowledge the partial support of the M.U.R.S.T., Italy, 
and the European Union for partial support.
N.S. wishes to thank the Noopolis-Sovena Foundation for financial support.
G.M. and N.S. thank CERN, Geneva, for ospitality during the final stage
of this work.



\begin{thebibliography}{10}

\bibitem{NOSTRO}
P.Marenzoni, G.Martinelli, N.Stella, M.Testa,
\newblock {\em Phys. Lett.}, {\bf B318}, (1993), 511.
 
\bibitem{PARRI}
C.~Bernard, C. Parrinello, A. Soni,
\newblock {\em Phys. Rev. } {\bf D49} (1994) 1585.


\bibitem{BP1}
N.~Brown and M.~Pennington.
\newblock {\em Phys. Rev.}, {\bf D38}, (1988), 2266.

\bibitem{BP2}
N.~Brown and M.~Pennington.
\newblock {\em Phys. Rev.}, {\bf D39}, (1989), 2723.

\bibitem{STINGL}
M.~Stingl.
\newblock {\em Phys. Rev.}, {\bf D34}, (1986), 3863.

\bibitem{GRIBOV}
V.N. Gribov.
\newblock {\em Nucl. Phys.}, {\bf B139}, (1978), 19.

\bibitem{ZW}
D.~Zwanziger.
\newblock {\em Nucl. Phys.}, {\bf B378}, (1992), 525.

\bibitem{MANDULA}
J.~E. Mandula and M. Ogilvie,
\newblock {\em Phys. Lett.}, {\bf B185}, (1987), 127.

\bibitem{GUPTA}
R.~Gupta {\em et~al}.
\newblock {\em Phys. Rev.}, {\bf D36}, (1987), 2813.



\bibitem{PARRIold}
C.~Bernard, C. Parrinello and A. Soni,
\newblock {\em Nucl. Phys. } \underbar{{\bf B (Proc. Suppl.) 30}} (1992) 535.


\bibitem{LHMC}.
P.~Marenzoni, G.~Pugnetti and P.~Rossi.
\newblock {\em Phys. Lett.}, {\bf B315} (1993) 152.


\bibitem{kn:Davies}
C.T.H. Davies.
\newblock {\em Phys. Rev.}, {\bf D37}, (1988), 1581.



\end{thebibliography}
\end{document}